\documentclass[12pt]{article}
\usepackage{graphicx,amsmath,bbm,mathrsfs,amssymb,psfrag}

\def\bmsigma{\boldsymbol{\sigma}}

\def\bmX{\boldsymbol{X}}

\def\bmC{\boldsymbol{C}}

\def\beq{\begin{equation}}
\def\eeq{\end{equation}}
\def\beqq{\begin{eqnarray}}
\def\eeqq{\end{eqnarray}}
\def\bmr{{\boldsymbol r}}
\def\bmell{{\boldsymbol \ell}}
\def\bmx{{\boldsymbol x}}
\def\bmy{{\boldsymbol y}}
\def\bmg{{\boldsymbol g}}
\def\bmw{{\boldsymbol w}}
\def\bmS{{\boldsymbol S}}
\def\bmD{{\boldsymbol D}}
\def\bmV{{\boldsymbol V}}
\def\bmOmega{{\boldsymbol \Omega}}
\def\llave{{\langle\!\langle}}
\def\ggave{{\rangle\!\rangle}}
\title{\bf Non-divisiblity and non-Markovianity in a Gaussian dissipative dynamics}

\author{Fabio Benatti$^{a,b}$, 
Roberto Floreanini$^{b}$, Stefano Olivares$^{c,a}$
\\
\small ${}^a$Dipartimento di Fisica, Universit\`a degli Studi di
Trieste, I-34151 Trieste, Italy,\\
\small ${}^b$Istituto Nazionale di Fisica Nucleare, Sezione di
  Trieste, I-34151 Trieste, Italy\\
\small ${}^c$Dipartimento di Fisica, Universit\`a degli Studi di Milano, 
	I-20133 Milano, Italy}

\date{\null}

\begin{document}

\maketitle

\begin{abstract}
\noindent
We study a stochastic Schr\"odinger equation that generates a family of Gaussian dynamical maps in one dimension permitting a detailed exam of two different definitions of non-Markovianity: one related to the explicit dependence of the generator on the starting time, the other to the non-divisibility of the time-evolution maps.
The model shows instances where one has non-Markovianity in both senses and cases when one has Markovianity in the second sense but not in the first one.
\end{abstract}

\vskip 1cm

Recent theoretical and experimental advances have aroused a lot of interest in non-Markovian effects when quantum systems interact with an environment which cannot be considered at equilibrium~\cite{Wilkie}-\cite{vas:PRA:11}.
More specifically, consider a system $S$ embedded in an environment $E$, under the hypothesis of an initial factorized state, i.e., a density matrix of the form $\rho\otimes\rho_E$; tracing away the environment degrees of freedom obtains an exact completely positive (CP) reduced dynamics for $S$ that sends an initial state $\rho$ at time $t_0\geq 0$ into a state $\rho_{t,t_0}$ at time $t\geq t_0$. This irreversible time-evolution is generated by an integro-differential equation of the form
\begin{equation}
\label{intkernel}
\partial_t\rho_{t,t_0}=\int_{t_0}^t{\rm d}u\, K_{t,u}[\rho_{u,t_0}]\ ,\quad\rho_{t_0,t_0}
=\rho\ ,
\end{equation}
where the operator kernel embodies  the dependence on the past history of the system.
The previous equation can be cast in the convolution-less form~\cite{ChruKoss1}
\begin{equation}
\label{TNL}
\partial_t\rho_{t,t_0}=\mathbb{L}_{t,t_0}[\rho_{t,t_0}]\ ,
\end{equation}
where the presence of memory effects is now incorporated in the dependence of the generator on the initial time $t_0$.
Because of this, the CP maps which solve~(\ref{TNL}),
\begin{equation}
\label{MCL1}
\Gamma_{t,t_0}=\mathcal{T}\exp\Bigg(\int_{t_0}^{t_1}{\rm d}u\,\mathbb{L}_{u,t_0}\Bigg)\ ,
\end{equation}
with $\mathcal{T}$ time-ordering, violate, in general, the (two-parameter) semigroup composition law, namely
\begin{equation}
\label{MCL}
\Gamma_{t,t_1}\circ\Gamma_{t_1,t_0}\neq\Gamma_{t,t_0}\ ,\qquad 0\leq t_0\leq t_1\leq t\ .
\end{equation}
Indeed, if $\mathbb{L}_{u,t_0}=\mathbb{L}_u$ then~(\ref{MCL1}) yields the equality in~(\ref{MCL}); vice versa, if in~(\ref{MCL}) the equality holds, by taking the time derivative of both sides with respect to $t$ one obtains $\mathbb{L}_{t,t_1}=\mathbb{L}_{t,t_0}$ for all $t_1\geq t_0\geq0$.
In~\cite{ChruKoss1}, the dependence of the generator $\mathbb{L}_{t,t_0}$ on $t_0$ and thus~(\ref{MCL})
is taken as a criterion  of non-Markovianity.
\par
On the other hand, in~\cite{WolfEisertCirac}--~\cite{LuWangSun} a different approach is considered whereby, given a one-parameter family of CP maps $\gamma_t$, $t\geq 0$, their non-Markovianity is related to non-divisibility, namely to the fact that no CP map $\Lambda_{t,u}$, $t\geq u\geq 0$, exists that connects the maps $\gamma_t$. In other words, the criterion of non-Markovianity becomes
\begin{equation}
\label{criterion2}
\gamma_t=\Lambda_{t,u}\circ\gamma_u\Longrightarrow \Lambda_{t,t_0}\quad\hbox{not CP}\ .
\end{equation}
If a CP $\Lambda_{t,u}$ existed, it would follow that certain CP monotone like the trace distance, the fidelity or the relative entropy should be decreasing: then, non-Markovianity is identified by the increase in time of such quantities which can also be taken as a measure of non-Markovianity.
\par
In order to study the two criteria of non-Markovianity,  we consider a stochastic Schr\"odinger equation originally proposed as a non-Markovian mechanism for the wave function collapse~\cite{BassiFerialdi}. Specifically,
we take a particle in one dimension subjected to a time-dependent random Hamiltonian of the form (for sake of simplicity, in the following, vector and matrix multiplication will be understood)
\begin{equation}
\label{stochHam}
\hat{H}^\bmw_t=\hat{H}-\,\bmw^T(t)\,\hat{\bmr}\ ,
\end{equation}
where the Hamiltonian $\hat{H}$ is at most quadratic in  position and momentum operators $\hat{\bmr}^T=(\hat{r}_1,\hat{r}_2)=(\hat{q},\hat{p})$, while $\bmw^T(t)=(w_1(t),w_2(t))$ is a Gaussian noise vector with zero mean and $2\times 2$ correlation matrix $\bmD(t,s)$:
\begin{equation}
\label{correlations}
\Big[\bmD(t,s)\Big]_{ij}=\llave w_i(t)\,w_j(s) \ggave,
\end{equation}
where $\llave\cdot\ggave$ denotes the average over the noise.
This latter matrix is real symmetric, $D_{ij}(t,s)=D_{ji}(s,t)$, and of positive-definite type, that is
\begin{equation}
\label{pos-def-type}
\sum_{i,j;t_a,t_b}\xi_i(t_a) \xi_j(t_b)\,D_{ij}(t_a,t_b)\geq 0\ ,\ \forall\ \xi(t_a)\in\mathbbm{R}^2\ ,
\end{equation}
for any choice of times $\{t_a\}_{a=1}^n$.
For each realization of the noise, the Schr\"odinger equation ($\hbar=1$)
\begin{equation}
\label{stocheq}
i\frac{{\rm d}\vert\psi^\bmw_t\rangle}{{\rm d}t}=
\left[\hat{H}-\bmw^T(t)\,\hat{\bmr}\right]
\vert\psi_t^\bmw\rangle\ ,
\end{equation}
generates unitary maps $\hat{U}^\bmw_{t,t_0}$ on the system Hilbert space that send an initial vector state $\vert\psi\rangle$ at time $t=t_0$ into $\vert\psi^\bmw_{t,t_0}\rangle$ at time $t$. Averaging the projector $\vert\psi^\bmw_{t,t_0}\rangle\langle\psi^\bmw_{t,t_0}\vert$ over the noise yields a density matrix
\begin{equation}
\label{densmat}
\rho_{t,t_0}= \llave\,\vert\psi^\bmw_{t,t_0}\rangle\langle \psi_{t,t_0}^\bmw\vert\ggave\ .
\end{equation}
In order to find $\hat{U}^\bmw_{t,t_0}$, one first goes to the interaction representation and sets:
\begin{eqnarray}
\nonumber
\vert\widetilde{\psi}^\bmw_{t,t_0}\rangle
&=&\hat{U}^\dag_{t-t_0}\vert\psi^\bmw_{t,t_0}\rangle\ ,\\
\label{2}
i\frac{{\rm d}\vert\widetilde{\psi}_{t,t_0}^\bmw\rangle}{{\rm d}t}&=& \bmw^T(t)\,\hat{\bmr}(t-t_0)\,
\vert\widetilde{\psi}_{t,t_0}^\bmw\rangle\ ,
 \end{eqnarray}
where $\hat{U}_t=\exp(-i\,\hat{H}\,t)$ and:
\begin{equation}
\label{sympl}
\hat{\bmr}(t)=\hat{U}_t^\dag\,\hat{\bmr}\,\hat{U}_t\equiv
\bmS_t\hat{\bmr}\ ,
\end{equation}
$\bmS_t$ being a suitable symplectic matrix.
For a given realization of the noise $\bmw(t)$, the solution is of the form $\vert\widetilde{\psi}^\bmw_{t,t_0}\rangle=
\widetilde{U}^\bmw_{t,t_0}\vert\psi\rangle$ where, a part for a pure phase,
\begin{eqnarray}
\label{3}
&&
\widetilde{U}^\bmw_{t,t_0}=
\exp\left\{-i\int_{t_0}^t{\rm d}u\,\hat{\bmw}^T(u)\,\hat{\bmr}(u-t_0)\right\}\\
&&
\vert\psi^\bmw_{t,t_0}\rangle=
\hat{U}_{t-t_0}\hat{U}^\bmw_{t,t_0}\,\vert\psi\rangle\ .
\end{eqnarray}
By averaging over the noise, the corresponding density matrix~(\ref{densmat}) satisfies:
$$
i\partial_t\rho_{t,t_0}=
\Big[\hat{H},\rho_{t,t_0}\Big]-\sum_{j=1}^2\Big[\hat{r}_j,
\llave w_j(t)\vert\psi^\bmw_{t,t_0}\rangle\langle \psi_{t,,t_0}^\bmw\vert\ggave\Big]\ .
$$
This stochastic Liouville equation can be turned into a standard master equation by means of the Furutsu-Novikov-Donsker relation~\cite{KonotopVasquez}:
\begin{equation}
\label{FND}
\llave \bmw(s) \bmX[\bmw]\ggave = \int_{-\infty}^{+\infty}\hskip-.5cm{\rm d}u\,\llave \bmw(s) \bmw(u)\ggave\,\llave\frac{\delta R[\bmw]}{\delta \bmw(u)}\ggave\ ,
\end{equation}
where $\bmX[\bmw]$ is a functional of the noise, $\delta/{\delta \bmw(u)}$ denotes the functional derivative with respect to the noise and $R[\bmw]$ is the density operator of the system.
With $R[\bmw]=\vert\psi^\bmw_{t,t_0}\rangle\langle \psi_{t,t_0}^\bmw\vert$, one gets:
\begin{equation}
\label{mastereq}
\partial_t\rho_{t,t_0}=
\mathbb{L}_{t,t_0}[\rho_{t,t_0}]=
-i\Big[\hat{H},\rho_{t,t_0}\Big]+
\mathbb{N}_{t,t_0}[\rho_{t,t_0}]
\end{equation}
with:
\begin{align}
\label{master2}
\mathbb{N}_{t,t_0}[\rho]&=\sum_{i,j=1}^2C_{ij}(t,t_0)\,\Big(\hat{r}_i\,\rho\,\hat{r}_j-
\frac{1}{2}\Big\{\hat{r}_j\hat{r}_i,\rho\Big\}\Big)\\
\bmC(t,t_0)&=\int_{t_0}^t{\rm d}u\, \Big[\bmD(t,u)\,\bmS_{u-t}+\bmS^T_{u-t}\bmD^T(t,u)\Big]\ .
\label{master2a}
\end{align}
If $\bmD(t,u) = \delta(t-u)\,\bmD$ (i.e., white noise) then one reduces to the Markovian Lindblad type dynamics with a time-independent positive Kossakowski matrix, namely $\bmC(t,t_0)=\bmD$~\cite{GKS,Lindblad}. In the time-dependent case, in order that the maps $\Gamma_{t,t_0}$ generated by $\mathbb{L}_{t,t_0}$ be CP, the Kossakowski matrix $\bmC(t,t_0)$ need not to be positive, as we explicitly show in the following. We shall seek a solution of~(\ref{mastereq}) in the form
\begin{equation}
\label{rsol}
\rho_{t,t_0}=\Gamma_{t,t_0}[\rho]=\int\frac{{\rm d}^2\bmr}{2\pi}\,G_{t,t_0}(\bmr)\,R(\bmr)\,\hat{W}(\bmS_{t-t_0}\bmr)\,,
\end{equation}
where we have introduced the Weyl operators:
\begin{equation}
\label{rot3}
\hat{W}(\bmr)={\rm e}^{i\,\bmr^T\bmOmega\,\hat{\bmr}}={\rm e}^{i(q\hat{p}-p\hat{q})}\ ,
\end{equation}
with $\bmr^T=(q,p)\in {\mathbbm R}^2$ and $\displaystyle
\bmOmega=\begin{pmatrix}0&1\cr-1&0\end{pmatrix}$,
and $R(\bmr)=\rm Tr[\rho\,\hat{W}(-\bmr)]$ is related to the initial condition by:
$$
\rho_{t_0,t_0}=\rho
=\int\frac{{\rm d}^2\bmr}{2\pi}\, R(\bmr)\,\hat{W}(\bmr)\,.
$$
Because the Hamiltonian $\hat{H}$ is at most quadratic and the matrix $S_t$ in~(\ref{sympl}) is symplectic, one finds:
$$
\hat{U}_t\,\hat{W}(\bmr)\,\hat{U}_t^\dag=\hat{W}(\bmS_t\bmr)\ .
$$
Direct insertion of~(\ref{rsol}) into~(\ref{mastereq}) yields
$$
\partial_tG_{t,t_0}(\bmr)=
-\Big[\bmr^T\,\bmS^T_{t-t_0}\bmC(t,t_0)\bmS_{t-t_0}\,\bmr\Big]\,G_{t,t_0}(\bmr)\, ,
$$
whence $G_{t,t_0}(\bmr)=\exp\left[-\frac{1}{2}\, \bmr^T\,\bmg(t,t_0)\,\bmr\right]$ with
\begin{eqnarray}
\label{exp1}
\bmg(t,t_0)&=&
2\,\int_{t_0}^t{\rm d}u\,\bmS^T_{u-t_0}\bmC(u,t_0)\bmS_{u-t_0}\\
\label{exp3}
&=&\int_{t_0}^t{\rm d}u\int_{t_0}^t\,{\rm d}v\,\bmS^T_{u-t_0}\bmD(u,v)\bmS_{v-t_0}\ .
\end{eqnarray}
Furthermore, since $\bmD(u,v)$ is of positive type, the matrix $\bmg(t,t_0)$ is positive definite and $G_{t,t_0}(\bmr)$ a real Gaussian function; the solution $\Gamma_{t,t_0}[\rho]$ can then  be cast in a continuous Kraus-Stinespring decomposition which guarantees the complete positivity of the maps $\Gamma_{t,t_0}$.
Let $G_{t,t_0}(\bmr)=\int_{\mathbb{R}^2}{\rm d}^2\bmx\, \delta(\bmx-\bmr)\,G_{t,t_0}(\bmx)$ with
$$
\delta(\bmx-\bmr)=\frac{1}{(2\pi)^2}\int_{\mathbb{R}^2}{\rm d}^2\bmy\,{\rm e}^{i\bmy^T\bmOmega(\bmx-\bmr)}\ .
$$
By inserting it into~(\ref{rsol}) and using
$\hat{W}(\bmx)\hat{W}(\bmr)\hat{W}^\dag(\bmx)={\rm e}^{-i\bmx^T\bmOmega \bmr}\hat{W}(\bmr)$,
one rewrites
\begin{equation}
\label{semig2a}
\Gamma_{t,t_0}[\rho]=\int_{\mathbb{R}^2}\frac{{\rm d}^2\bmy}{2\pi}F_{t,t_0}(\bmy)\, \hat{U}_{t-t_0}\hat{W}(\bmx)\,\rho\,\hat{W}^\dag(\bmx)\hat{U}^\dag_{t-t_0}\\
\end{equation}
with the Fourier transform
\begin{equation}
\label{semig2b}
F_{t,t_0}(\bmy)=\int_{\mathbb{R}^2}\frac{{\rm d}^2\bmx}{2\pi}\,{\rm e}^{i\bmy^T\bmOmega \bmx}G_{t,t_0}(\bmx)\ ,
\end{equation}
also a real Gaussian, hence a positive function.
\par
Using~(\ref{rsol}) one can study the composition properties of the maps $\Gamma_{t,t_0}$; since:
\begin{equation}\label{comp-law}
\Gamma_{t_2,t_1}\circ\Gamma_{t_1,t_0}[\rho]=
\int\frac{{\rm d}^2\bmr}{2\pi}\, G_{t_2,t_1}(\bmS_{t_1-t_0}\bmr)\,G_{t_1,t_0}(\bmr)
\,R(\bmr)\,\hat{W}(\bmS_{t_2-t_0}\bmr)\nonumber ,
\end{equation}
in order to to satisfy the semigroup composition law $\Gamma_{t_2,t_1}\circ\Gamma_{t_1,t_0}=\Gamma_{t_1,t_0}$
one should have
$$
G_{t_2,t_1}(\bmS_{t_1-t_0}\bmr)\,G_{t_1,t_0}(\bmr)=G_{t_2,t_0}(\bmr)\,.
$$
Using~(\ref{exp3}), one instead finds that
\begin{eqnarray}
&&\nonumber
\Bigg(\int_{t_1}^{t_2}\int_{t_1}^{t_2}+\int_{t_0}^{t_1}\int_{t_0}^{t_1}{\rm d}u\,{\rm d}v\Bigg)
\Bigg(\bmS^T_{u-t_0}\bmD(u,v)\bmS_{v-t_0}\Bigg)\\
\label{neq}
&&\hskip 5cm\neq
\int_{t_0}^{t_2}{\rm d}u\int_{t_0}^{t_2}{\rm d}v\,\bmS^T_{u-t_0}\bmD(u,v)\bmS_{v-t_0}\ .
\end{eqnarray}
This fact remains true even when $\bmD(s,u)=\bmD(|s-u|)$ in which case from~(\ref{exp3}) we have
$$
\bmg(t,t_0)=\int_0^{t-t_0}{\rm d}u\int_0^{t-t_0}\,{\rm d}v\,\bmS^T_u\bmD(u,v)\bmS_v
$$
and
$\Gamma_{t,t_0}=\Gamma_{t-t_0,0}$.
\par
Consider the master equation~(\ref{mastereq}); if $t_0=0$ its solutions $\rho_{t,0}=\Gamma_{t,0}[\rho]$ propagate the initial state $\rho$ from $t_0=0$ to $t\geq 0$. Because of the above result,  $\Gamma_{t,0}\neq\Gamma_{t,t_0}\circ\Gamma_{t_0,0}$. However, setting $t_0=0$ in~(\ref{mastereq}) and searching a solution $\Lambda_{t,t_0}[\rho]$ in the form~(\ref{rsol}), one gets
\begin{equation}
\label{Lambda-time0}
\Lambda_{t,t_0}[\rho]=\int\frac{{\rm d}^2\bmr}{2\pi}\,L_{t,t_0}(\bmr)\,R(\bmr)\,\hat{W}(\bmS_{t-t_0}\bmr)\\
\end{equation}
where $L_{t,t_0}(\bmr)=\exp\left\{-\frac{1}{2} \bmr^T \bmell(t,t_0) \bmr\right\}$ with:
\begin{eqnarray}
\label{Lambda-time2}
\bmell(t,t_0)&=&\int_{t_0}^t{\rm d}u\, \bmS^T_{u-t_0}\,\bmC(u,0)\,\bmS_{u-t_0}\\
\label{Lambda-time3}
&=&
\int_{t_0}^t{\rm d}u\int_0^u{\rm d}v\, \bmS^T_{u-t_0}\,\bmD(u,v)\,\bmS_{v-t_0}\,.
\end{eqnarray}
The function $L_{t,t_0}(r)$ plays the role of $G_{t,t_0}(\bmr)$ in~(\ref{rsol}) to which it reduces when $t_0=0$; that is $\Lambda_{t,0}=\Gamma_{t,0}$.
Note however that, in contrast to $\bmg(t,t_0)$ in~(\ref{exp1}), in $\bmell(t,t_0)$ one integrates $\bmC(u,0)$, not $\bmC(u,t_0)$, from $t_0$ to $t$. As a consequence, $\Gamma_{t,0}=\Lambda_{t,t_0}\circ\Gamma_{t_0,0}$; indeed,
\begin{equation}
\Lambda_{t,t_0}\circ\Gamma_{t_0,0}[\rho]=
\int_{\mathbb{R}^2}\frac{{\rm d}^2\bmr}{2\pi}\,
L_{t,t_0}(\bmS_{t_0}\bmr)\,L_{t_0,0}(\bmr)\,R(\bmr)\,
\hat{W}(\bmS_t\bmr)\ ,\nonumber
\end{equation}
where now, unlike in~(\ref{neq}),
\begin{align}
\bmS^T_{t_0}\,\ell(t,t_0)\,\bmS_{t_0}+\bmell(t_0,0)&=
\Bigg(\int_{t_0}^t\int_0^u +\int_0^{t_0}\int_0^u\Bigg){\rm d}u\,{\rm d}v\,\bmS^T_u\,\bmD(u,v)\,\bmS_v\nonumber\\
&=
\int_0^t{\rm d}u\int_0^u{\rm d}v\,\bmS^T_u\,\bmD(u,v)\,\bmS_v=\bmell(t,0)\ .\nonumber
\end{align}
However, contrary to the maps $\Lambda_{t,0}=\Gamma_{t,0}$ which, as we have seen, are CP, the maps
$\Lambda_{t,t_0}$ cannot be CP as this would imply~\cite{LainePiiloBreuer} the positive definiteness of the matrix $\bmC(t,t_0)$ in~(\ref{master2}).
In fact, the maps $\Lambda_{t,t_0}$ are in general not even positive.
\par
All these various possibilities can be seen in a concrete example; consider a free particle of unit mass, $\hat{H}=\hat{p}^2/2$,
so that $\bmS_t=\begin{pmatrix}1&t\cr0&1\end{pmatrix}$, and a diagonal noise with correlation matrix given by
\begin{equation}
\label{corrmat}
\bmD(t,u)=\frac{\gamma\,{\rm e}^{-\gamma|t-u|}}{2}\begin{pmatrix}d_q&0\cr0&d_p\end{pmatrix}\ .
\end{equation}
First suppose the noise couples only to the position operator: $d_q=1$, $d_p=0$; then,
from~(\ref{master2a}),
\begin{equation}
\label{CM0}
\bmC(t,t_0)=\begin{pmatrix}
1-{\rm e}^{-\gamma (t-t_0)}&\frac{{\rm e}^{-\gamma (t-t_0)}[1+\gamma (t-t_0)]-1}{2\gamma}\cr
\frac{{\rm e}^{-\gamma (t-t_0)}[1+\gamma(t-t_0)]-1}{2\gamma}&0\end{pmatrix}
\end{equation}
has a negative eigenvalue for all $t>t_0\geq 0$. In spite of the non-positivity of the Kossakowski matrix in~(\ref{master2a}), the maps $\Gamma_{t,t_0}$ in~(\ref{semig2a}) are nevertheless CP for all $0\leq t_0\leq t$.
\par
We consider, as initial condition at $t_0$, a Gaussian state $\rho_{\bmsigma}$ with covariance matrix (CM) $\bmsigma$ and zero first moments,
${\rm Tr}\left[\rho_{\bmsigma}\,\hat{W}(-\bmr)\right]=\exp\left\{-\frac{1}{2} \bmr^T\,(\bmOmega\,\bmsigma\,\bmOmega^T)\,\bmr\right\}$.
Using~(\ref{rsol}), $\Gamma_{t,t_0}$ maps $\rho_{\bmsigma}$ to the Gaussian state
${\rm Tr}\left[\Gamma_{t,t_0}[\rho_{\bmsigma}]\hat{W}(\bmr)\right]= \exp\left\{-\frac{1}{2} \bmr^T\bmOmega^T\bmsigma_{t,t_0}\bmOmega \bmr\right\}$, where $\bmsigma_{t,t_0}=\bmS_{t-t_0}\bmsigma \bmS^T_{t-t_0}+\widetilde{\bmg}(t,t_0)$ with
\begin{equation}
\label{NT}
\widetilde{\bmg}(t,t_0)=
\int_{t_0}^t{\rm d}u\, \bmOmega^T\,\bmS^T_{u-t}\,\bmC(u,t_0)\,\bmS_{u-t}\,\bmOmega\ .
\end{equation}
Instead, if the same initial condition is taken for the maps $\Lambda_{t,t_0}$, the matrix $\widetilde{\bmg}(t,t_0)$ is to be substituted by
\begin{equation}
\label{newmat}
\widetilde{\bmell}(t,t_0)
=\int_{t_0}^t{\rm d}u\,\bmOmega^T\, \bmS^T_{u-t}\,\bmC(u,0)\,\bmS_{u-t}\, \bmOmega\ .
\end{equation}
If we choose $\bmsigma=\bmS_{t_0-t}\bmsigma_0\bmS^T_{t_0-t}$ and expand $\bmsigma_{t,t_0}=\bmsigma_0+\widetilde{\bmell}(t,t_0)$ to first order about $t_0$, we have:
\begin{align}
\bmsigma_{t,t_0}&\simeq\bmsigma_0+
(t-t_0)\bmOmega^T\,\bmC(t_0,0)\,\bmOmega\,,
\end{align}
where $\bmC(t_0,0)$ is calculated from Eq.~(\ref{CM0}). Now,
the second matrix at the l.h.s.~is real symmetric and has one positive and one negative eigenvlaue, $\lambda\geq 0$ and $-\mu<0$; let $\bmV$ be the symplectic, orthogonal matrix which diagonalizes it. Then,
choosing an initial state with CM diagonal in the same basis, i.e.,  $\bmsigma_0 = {\rm Diag}[\sigma_{qq},\sigma_{pp}]$, such that $\bmsigma_{0} + \frac{i}{2}\bmOmega \ge 0$ (positivity of the initial state), one gets:
$$
\bmsigma_{t,t_0}\simeq \bmV^T\,\begin{pmatrix}
\sigma_{qq}+\lambda (t-t_0)&0\cr0&\sigma_{pp}-\mu (t-t_0)
\end{pmatrix}\, \bmV\ ,
$$
and a sufficiently small $\sigma_{pp}$ would yield a non positive-definite CM $\sigma_{t,t_0}$, thus exhibiting the non-positivity of the map $\Lambda_{t,t_0}$.  The non-positive preserving character of $\Lambda_{t,t_0}$ is exposed by very specific states; on other states as, for instance, on all those of the form $\Gamma_{t_0,0}[\rho]$ it acts perfectly well for $\Lambda_{t,t_0}\circ\Gamma_{t_0,0}=\Gamma_{t,0}$.  In addition, starting from $t_0=0$, $\Lambda_{t,0}=\Gamma_{t,0}$ is CP.
\par
Therefore, in this case the master equation~(\ref{mastereq}) generates a non-Markovian dynamics both according to the criterion~(\ref{MCL}), since the generator $\mathbb{L}_{t,t_0}$ depends on the initial time $t_0$ and also according to the other criterion~(\ref{criterion2}).
In fact, the family of maps $\Gamma_{t,0}$ is non-divisible for $\Lambda_{t,t_0}$ is uniquely defined and non-positive.
\par
\begin{figure}[h]
\begin{center}
\includegraphics[scale=1.6]{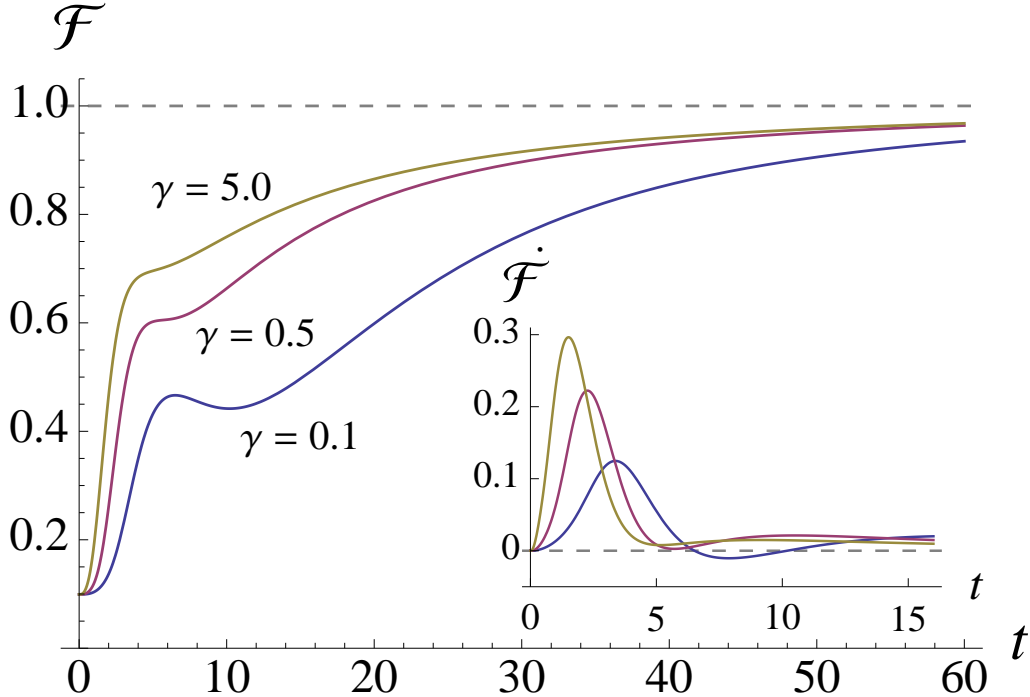}
\caption{\label{f:fid}\small Plot of the time evolution of the fidelity ${\cal F}$ between two Gaussian states $\rho_k$, $k=1,2$, with zero first moments and CMs $\bmsigma_k=\frac12 {\rm Diag}\left[\exp(2r_k),\exp(-2r_k)\right]$, with $r_1=-r_2=1.5$, evolving under the map $\Gamma_{t,0}$ for different values of $\gamma$. The inset refers to the time derivative of the fidelity. The non-monotonic behavior denotes non-Markovian evolution \cite{vas:PRA:11}; note that as $\gamma$ increases, ${\cal F}$ becomes monotonic.}
\end{center}
\end{figure}
Since $\Lambda_{t,t_0}$ is not (completely) positive, certain quantities that exhibit monotonic behavior under CP maps fail to do so when evolving the system from time $t_0$ to time $t$.
One of such quantities is the fidelity~\cite{vas:PRA:11} $\mathcal{F}(t)=\mathcal{F}(\Gamma_{t,0}[\rho_1],\Gamma_{t,0}[\rho_2])$ of two states $\rho_1$ and $\rho_2$ evolving in time according to $\Gamma_{t,0}$. While $\mathcal{F}(t)\geq \mathcal{F}(0)$ for all $t\geq 0$, $\mathcal{F}(t_0+t)$ may become smaller than $\mathcal{F}(t_0)$ for some $t,t_0>0$. This is showed in Fig.~\ref{f:fid} for two Gaussian states with zero first moments and ``squeezed'' CM. As one may expect, the effect disappears when $\gamma$ increases towards the Markovian limit.
\par
On the other hand, if in~(\ref{stocheq}), the noise affects the particle momentum only, namely if $d_q=0$, $d_p=1$, then, from~(\ref{corrmat}),
\begin{equation}
\label{stochmom}
C(t,t_0)
=(1-{\rm e}^{-\gamma(t-t_0)})\,\begin{pmatrix}
0&0\cr0&1\end{pmatrix}
\end{equation}
is positive definite. It follows that the intertwining map $\Lambda_{t,t_0}$ is CP, whence the family of maps $\Gamma_{t,0}$ is divisible and Markovian according to the criterion~(\ref{criterion2}). However,  it is non-Markovian according to the other criterion~(\ref{MCL}). Indeed, the generator  resulting from~(\ref{stocheq}) depends on the starting time $t_0$.
\par
In conclusion, the analysis of above examples indicates that the criterion identifying non-Markovianity with the explicit dependence of the generator ${\mathbb L}_{t,t_0}$ on the starting time $t_0$ appears stronger than the criterion based on the non-divisibility of the maps $\Gamma_{t,0}$.  Indeed, on one hand, we have provided a case where the map $\Gamma_{t,0}$ is divisible, yet the generator of $\Gamma_{t,t_0}$ explicitly depends on the initial time $t_0$; on the other hand, a Markovian evolution according to the first criterion readily implies the semigroup composition law, {\it i.e.}, (4) with the equality sign, hence divisibility of $\Gamma_{t,0}$.  Nevertheless, the non-divisibility criterion is the only one at disposal when one is presented just with the family of maps $\Gamma_{t,0}$: in such a case, one may reconstruct the generator ${\mathbb L}_{t,0}$ starting from $t_0 = 0$, but, in general, no information is available on the full generator ${\mathbb L}_{t,t_0}$ at $t_0>0$.

\vskip .5cm
\noindent
{\bf Acknowledgments} FB and RF thank A.~Bassi and L.~Ferialdi for useful discussions. SO acknowledges useful discussions with M.~G.~A.~Paris and R.~Vasile and financial support from MIUR (FIRB ``LiCHIS'' - RBFR10YQ3H) and from the University of Trieste (``FRA 2009'').

\end{document}